\documentstyle[12pt]{article}
\input tcilatex

\begin{document}

\author{M. Lisowski and E. Zipper  \\ 
Institute of Physics, University of Silesia\\ul. Uniwersytecka 4, 40-007
Katowice, Poland}
\title{Orbital Magnetic Ordering in Disordered Mesoscopic Systems}
\date{April 28, 1998}
\maketitle

\begin{abstract}
We present some model calculations of persistent currents in disordered one-
and two-dimensional mesoscopic systems. We use the tight-binding model and
calculate numerically the currents in small systems for several values of
disorder. Next we fit appropriate analytical formulae, and using them we
find self-sustaining currents and critical fields in larger, more realistic
systems with different shapes of the Fermi surfaces (FS).

\ 


\ 

\end{abstract}

\baselineskip=25pt

\newpage\ 

\section{Introduction}

With advances in technology, fabrication of submicron devices has become
possible. Such systems exhibit quantum coherence which is the subject of
extensive experimental and theoretical studies.

Recently, in a series of papers [1], model considerations of self-sustaining
persistent currents in a set of clean mesoscopic metallic rings have been
presented. In this paper we investigate whether such currents survive in the
presence of scattering driven by impurities and temperature. We will
consider a set of concentric mesoscopic rings stacked along a certain axis
and a set of concentric two-dimensional (2-D) cylinders. Such samples can be
obtained by lithographic methods from metals or semiconductors \cite
{lithograph}.

\section{Quasi One-dimensional Mesoscopic Rings}

Let us consider a set of quasi one-dimensional (1-D) metallic mesoscopic
rings stacked along $y$ axis threaded by the external magnetic flux $\phi _e$%
. We assume that each ring posseses the same number of conducting electrons $%
N_e$ (even or odd). The situation when some of the rings carry an odd number
and other carry an even number of electrons has been considered in \cite
{Wohlleben}. The long range magnetostatic (current-current) interaction
between electrons from different rings is taken into account in the mean
field approximation (MFA).

The gauge-invariant tight-binding Hamiltonian for a single ring is of the
form: 
\begin{equation}
\label{1imp}\widehat{H}_1=\stackunder{n=1}{\stackrel{N}{\sum }}%
[(2t+V_n)c_n^{+}c_n-te^{i\theta _{n,n+1}}c_{n+1}^{+}c_n-te^{-i\theta
_{n,n+1}}c_n^{+}c_{n+1}], 
\end{equation}
where $t=\hbar ^2/(2m_ea^2)$ is the hopping matrix element, $a$ is the
lattice constant; $c_n^{+},c_n$ are the creation and annihilation operators; 
$N$ is the number of sites in each ring (channel); $\theta _{n,n+1}$ comes
from the magnetic flux $\phi :$%
\begin{equation}
\label{2imp}\theta _{n,n+1}=\frac e\hbar \stackrel{r_{n+1}}{\stackunder{r_n}{%
\int }}\underline{A}\cdot d\underline{l}=\frac{2\pi }N\frac \phi {\phi
_0}\equiv \theta , 
\end{equation}

where \underline{$A$} is the vector potential, and $\phi _0=h/e.$

The flux $\phi $ is composed of two parts, 
\begin{equation}
\label{3imp}\phi =\phi _e+\phi _I,\quad \phi _I={\cal L}I(\phi ), 
\end{equation}
i.e., each electron moves in the external magnetic flux $\phi _e$ and in the
flux coming from the total current $I$ in the system; ${\cal L}$ is the
selfinductance coefficient.

The disorder is given by a random choice of the on-site potentials $V_n$
from a rectangular distribution of width $W=\kappa t,\ \kappa \geq 0$, and
strength $-W/2$ to $W/2$.

In the case of a clean system ($V_n=0$) we can calculate the current $I$ in
the system, diagonalizing the Hamiltonian (\ref{1imp}) directly, by using
the Fourier transform. We get%
$$
I(\phi )=M_tI_1(\phi ), 
$$
\begin{equation}
\label{4imp}I_1(\phi )=-\stackunder{\alpha }{\sum }\frac{\partial {\cal E}%
_\alpha }{\partial \phi }=\frac{e\hbar }{Nm_ea^2}\stackunder{\alpha }{\sum }%
f_\alpha \sin \left[ \frac{2\pi }N\left( \alpha -\frac \phi {\phi _0}\right)
\right] , 
\end{equation}
where ${\cal E}_\alpha =2t\left[ 1-\cos \frac{2\pi }N\left( \alpha -\frac
\phi {\phi _0}\right) \right] ,$ $f_\alpha $ is the Fermi-Dirac distribution
function, $\alpha $ is the orbital quantum number for an electron going
around the ring ($\alpha =0,\pm 1,\pm 2,...$); $M_t=MP$, $M$ and $P$ are the
numbers of rings along and perpendicular to $y$ axis respectively $(P\ll N)$.

In this paper we concentrate on a disordered system ($V_n\neq 0$). The
Hamiltonian (\ref{1imp}) can be written in the form of a matrix ($N$x$N$): 
\begin{equation}
\label{7imp}\widehat{H}_1=\left( 
\begin{array}{cccccc}
2t+V_1 & -te^{-i\theta _{1,2}} &  &  &  & -te^{i\theta _{N,1}} \\ 
-te^{i\theta _{1,2}} & 2t+V_2 & -te^{-i\theta _{2,3}} &  & 0 &  \\  
& -te^{i\theta _{2,3}} & 2t+V_3 & -te^{-i\theta _{3,4}} &  &  \\  
&  & ... & ... & ... &  \\  
& 0 &  & -te^{i\theta _{N-2,N-1}} & 2t+V_{N-1} & -te^{-i\theta _{N-1,N}} \\ 
-te^{-i\theta _{N,1}} &  &  &  & -te^{i\theta _{N-1,N}} & 2t+V_N 
\end{array}
\right) 
\end{equation}
where the base vectors are chosen as follows:%
$$
\begin{array}{c}
c_1^{+}\left| 0\right\rangle =\left| 100...0\right\rangle \equiv \left|
1\right\rangle \\ 
c_2^{+}\left| 0\right\rangle =\left| 010...0\right\rangle \equiv \left|
2\right\rangle \\ 
..... \\ 
c_N^{+}\left| 0\right\rangle =\left| 000...1\right\rangle \equiv \left|
N\right\rangle 
\end{array}
$$
$\left| 0\right\rangle =\left| 000...0\right\rangle $ - the vacuum state.

The above matrix we have diagonalized numerically for different $N$ (%
\TEXTsymbol{<}100) and different $W/t$ using the ''Monte Carlo'' method \cite
{Koonin}. We also calculated the current $I_1(\phi ,W/t,T)$ by numerical
differentiation; it is presented in Fig. 1 for $N=80$ by symbols \cite
{LisZipKos}. We see that the impurity scattering leads to the amplitude
reduction and the shift of the maximum of the $I_1(\phi )$ characteristics.
We note \cite{IBM} that elastic scattering has a very similar effect on the $%
I_1(\phi )$ as temperature. Whereas a non-zero temperature leads to a
redistribution of electrons among the energy levels, collisions lead to a
change in the levels themselves. If the average collision time is $\tau $,
then the uncertainty in the electron energy is of the order of $\hbar /\tau
, $ inverse proportional to the elastic scattering mean free path $l_e\sim
(t/W)^2$ \cite{W/t} (the localization length $\xi $ in one dimension). If $%
\tau $ is not too short, it plays a similar role as the temperature \cite
{Abrikosov}.

Because of the limited capacity of computers and long time needed for
computing, it is very difficult to calculate in this way the current $%
I_1(\phi ,W/t,T)$ for $N>100.$ However we want to investigate the
possibility of the existance of self-sustaining current in the system with
impurities. The phenomenon of such current is a collective effect \cite
{SzopZip} and can be obtained only if the number of interacting electrons is
relatively large, i.e. for large (but still mesoscopic) systems. The
selfconsistency formula for the current at $\phi _e=0$, obtained by the use
of Eq. (\ref{3imp}), is of the form: 
\begin{equation}
\label{self}I(\phi ,W/t,T)=\frac \phi {{\cal L}}, 
\end{equation}
where 
\begin{equation}
\label{10imp}{\cal L}=\frac{\mu _0\pi R^2}{l^2}\left( \sqrt{l^2-R^2}%
-R\right) , 
\end{equation}
$l=Mb$ is the length of the cylinder made of a set of mesoscopic rings, $b$
is the distance between rings; $\mu _0$ is the magnetic permeability of free
space.

To get a stable, nontrivial solution of Eq. (\ref{self}) we need to consider
a set of mesoscopic rings with $N\gg 100$. Even though exact numerical
solutions of the current in the presence of impurities were performed for
very small samples, they show generic features for all mesoscopic rings \cite
{IBM}. Thus we have fitted an analytical formula to the numerical results of
Fig. 1 and we assumed that it can be used for the qualitative analysis of
Eq. (\ref{self}) for larger rings. $I_1$ is periodic in $\phi /\phi _0$,
with period 1, and can be expressed as a Fourier sum \cite{Gefen}, \cite
{Riedel}. The best fitting, denoted in Fig. 1 by solid line, we have
obtained for the formula: 
\begin{equation}
\label{9imp}
\begin{array}{c}
I_1(\phi ,\gamma ,T)=
\stackrel{\infty }{\stackunder{q=1}{\dsum }}\dfrac{4I_0}\pi \left( \dfrac
L{2\gamma }+\dfrac T{T^{*}}\right) \dfrac{\exp \left[ -q\left( \frac L\gamma
+\frac T{T^{*}}\right) \right] }{1-\exp \left[ -q\left( \frac L\gamma
+2\frac T{T^{*}}\right) \right] }\times  \\ \times \cos (qk_FL)\sin \left(
2\pi q\dfrac \phi {\phi _0}\right) ,
\end{array}
\end{equation}
where $I_0=e\hbar \sin (k_Fa)/(Nm_ea^2)$, $k_F=N_e\pi /L$, $L=Na$ is the
circumference of a ring; $T^{*}=\Delta /(2\pi ^2k_B)$ is a characteristic
temperature that separates the high- and low-temperature regimes, $\Delta
=4\pi t\sin (k_Fa)/N$ is the level spacing at zero flux at the Fermi surface
(FS), $k_B$ is the Boltzmann constant; $1/\gamma $ is a disorder parameter, $%
\gamma \cong 10a(t/W)^{3/2}\ln N$.

To account for different values of $\gamma $ ($W/t$) in a set of $M_t$
mesoscopic rings we have performed a quenched average of the current $I(\phi
,\gamma ,T)$, where disorder was given by a Gaussian distribution with
standard deviation $\Delta \gamma =\pm 10\%\overline{\gamma }$, $\overline{%
\gamma }$ is a mean value of $\gamma $. We have obtained a very small
decrease in the amplitude of $[I(\phi ,\gamma ,T)]$, where [...] denotes the
quenched average, comparing to the case with the same disorder in each ring.

In the clean system at $T=0$ $K$ we would have \cite{Gefen}: 
\begin{equation}
\label{5imp}I(\phi )=M_t\stackrel{\infty }{\stackunder{q=1}{\sum }}\frac{2I_0%
}{\pi q}\cos (qk_FL)\sin \left( 2\pi q\frac \phi {\phi _0}\right) . 
\end{equation}

Now we are in position to look for the self-sustaining currents. They can be
obtained from the self-consistent equation for the total current (\ref{self}%
), where the current $I=M_tI_1$ is taken with the quenched average, $I_1$ is
given by the formula (\ref{9imp}).

The case of a clean system, in the free electron model, has been analysed in 
\cite{SzopZip}. Here we extend earlier studies to include the dependences on
disorder and temperature in the tight-binding model with the half-filled
band. The results depend on the number of electrons in a single ring.

In the case of an even number of electrons in each ring the current is
paramagnetic and the stable solutions of Eq. (\ref{self}) being the
intersections of the curves marked by circle correspond to a spontaneous
current $I_s$. They are presented in Fig. 2 for different parameters $%
\overline{\gamma }$. In Fig. 3 we also present the temperature dependence of
persistent current and we see that both temperature and disorder act almost
in the same way.

In the case of an odd number of electrons in each ring the current is
diamagnetic for small $\phi $ and the stable solutions of Eq. (\ref{self})
presented in Fig. 4 correspond to flux trapping, the phenomenon known in
superconductors.

The temperature at which the transition to the state with the
self-sustaining current occurs is denoted by $T_c$. We see that $T_c$
decreases with increasing the impurity content and that flux trapping can be
obtained only in weakly disordered systems.

\section{Two-dimensional Mesoscopic Cylinders}

Let us consider now a set of coaxial closely packed two-dimensional (2-D)
cylinders \cite{Efros}. The appropriate tight-binding Hamiltonian for a
single cylinder is of the form: 
\begin{equation}
\label{11imp}
\begin{array}{c}
\widehat{H}=\stackrel{M}{\stackunder{m=1}{\dsum }}\stackrel{N}{\stackunder{%
n=1}{\dsum }}[(2t+V_{nm})c_{nm}^{+}c_{nm}-te^{i\theta
_{n,n+1}}c_{n+1m}^{+}c_{nm}+ \\  \\ -te^{-i\theta
_{n,n+1}}c_{nm}^{+}c_{n+1m}-t_bc_{nm+1}^{+}c_{nm}-t_bc_{nm}^{+}c_{nm+1}], 
\end{array}
\end{equation}
where $t_b=\hbar ^2/(2m_eb^2)$ is the hopping element in the $y$ direction, $%
b$ is the lattice constant in the $y$ direction.

In this case to get the current $I$ in a disordered system, we have
performed similar procedures to the previous 1-D case. Finally we have
obtained: 
\begin{equation}
\label{13imp}
\begin{array}{c}
I(\phi ,\gamma ,T)=P 
\stackrel{M}{\stackunder{m=1}{\dsum }}\stackrel{\infty }{\stackunder{q=1}{%
\dsum }}\dfrac{4I_0(m)}\pi \left( \dfrac L{2\gamma }+\dfrac T{T^{*}}\right) 
\dfrac{\exp \left[ -q\left( \frac L\gamma +\frac T{T^{*}}\right) \right] }{%
1-\exp \left[ -q\left( \frac L\gamma +2\frac T{T^{*}}\right) \right] }\times
\\ \times \cos [qk_{F_x}(m)L]\sin \left( 2\pi q\dfrac \phi {\phi _0}\right)
, 
\end{array}
\end{equation}
where $1/\gamma $ is a disorder parameter, $\gamma \sim (t/W)^{3/2}\ln N$;

$I_0(m)=e\hbar \sin [k_{F_x}(m)a]/(Nm_ea^2)$, $k_{F_x}(m)$ is the radius of
the FS corresponding to the channel $m$ which depends on the shape of the FS.

After performing a quenched average of $I(\phi ,\gamma ,T)$ over $\gamma $ ($%
W/t$) with the Gaussian distribution we can look for self-sustaining
currents.

In the clean system at $T=0$ $K$ we would have again \cite{IBM}: 
\begin{equation}
\label{2Dclean}I(\phi )=P\stackrel{M}{\stackunder{m=1}{\sum }}\stackrel{%
\infty }{\stackunder{q=1}{\sum }}\frac{2I_0(m)}{\pi q}\cos
(qk_{F_x}(m)L)\sin \left( 2\pi q\frac \phi {\phi _0}\right) . 
\end{equation}

The self-sustaining currents can be obtained from equations (\ref{self}) and
(\ref{13imp}). It is known that for 2-D systems and for 3-D systems with 2-D
conduction (an example of such structures are high $T_c$ superconductors in
a normal state \cite{StebSzopZip}) persistent currents depend crucially on
the strength of the phase correlation between currents of different channels 
\cite{IBM}, \cite{StebSzopZip} and therefore we have analysed three
different cases namely the systems with rectangular, triangular and
half-circular FS. The $k_{F_x}(m)$ is then given by 
\begin{equation}
\label{15cimp}k_{F_x}(m)= 
\begin{array}{l}
\pm \frac \pi a 
\text{ for }|k_{F_y}(m)|\leq \frac \pi b, \\ \ 0\text{\quad for other }%
k_{F_y}(m), 
\end{array}
\end{equation}
\begin{equation}
\label{15aimp}k_{F_x}(m)=\pm \frac 1a\arccos \left[ -\frac{a^2}{b^2}\cos
(k_{F_y}(m)b)\right] , 
\end{equation}
and 
\begin{equation}
\label{15bimp}k_{F_x}(m)=\pm \sqrt{\frac 2{a^2}+\frac 2{b^2}-k_{F_y}^2(m)} 
\end{equation}
respectively; $k_{F_y}(m)=m\pi /((M+1)b).$ Such shapes of the FS can be
obtained in the tight-binding approximation depending on the crystal
symmetry and on the band filling.

The most favorable situation is for the rectangular FS (found in bcc
crystals for nearly half filled band) where self-sustaining currents $I_s$
are the easiest to obtain and are the largest. In the second case (crystals
with simple cubic symmetry at half filling), it is possible to obtain
self-sustaining currents only for weakly disordered systems (ballistic
regime) and for the maximal interchannel phase correlation, where $L/l\cong
2+h/3$, $h$ is a positive integer; Fig. 5. In the third case (found at very
low band filling), we have not obtained such currents, for physical values
of parameters, because the interchannel correlations are too small \cite
{StebSzopZip}. The $[I(\phi )]$ characteristics for different shapes of the
FS and for $\overline{\gamma }$ = 50000 \AA\ are presented in Fig. 6.

Self-sustaining currents are a hallmark of a phase coherence. Thus we can
introduce, in analogy to superconductors, a notion of a critical field $H_c$
below which the system is in a coherent state. The system is in a coherent
state if we get at $\phi _e=0$ the nonzero crossing of the currents $I(\phi
) $ and $\frac \phi {{\cal L}}$ (Eq. (\ref{self})). Let us denote by $\phi
_{\max }$ the value of $\phi $ at which $I(\phi )$ reaches its maximum value 
$I_{\max }\equiv I(\phi _{\max })$.

If 
\begin{equation}
\label{max}I(\phi _{\max })\geq \frac{\phi _{\max }}{{\cal L}}, 
\end{equation}
we get a non-zero solution. Accepting the interpretation \cite{Bloch} that
the critical field $H_c$ in the cylinder is determined by the maximum value
of the current $I_{\max }$ one finds from Eq. (\ref{max}): 
\begin{equation}
\label{Hc}H_c=\frac{{\cal L}I_{\max }}{\mu _0\pi R^2}. 
\end{equation}
The magnitude of $I_{\max }$ depends on temperature, disorder and on the
geometry of the FS and can be extracted from Figs. 4, 5.

The critical field $H_c$ as a function of the disorder parameter $1/\gamma $
and temperature $T$ is presented in Figs. 7 and 8 respectively. We see that
it decreases with the increasing disorder and likewise decreases with the
increasing temperature.

\section{Conclusions}

In the presented paper we have addressed the question of whether
self-sustaining currents in nonsuperconducting rings and cylinders survive
in the presence of disorder and finite temperature. The answer to this
question is positive. We have performed some model considerations which show
that self-sustaining currents can be obtained in weakly disordered systems
at temperatures of the order of 1 $K$ in the systems with strong
correlations among the channel currents, i.e. in the systems with the FS
having flat regions.

From the presented $I(\phi )$ characteristics we were able to determine the
critical field $H_c$, below which the system is in a phase coherent state.
We have shown that $H_c$ decreases with increasing disorder and temperature
respectively.

\section{Acknowledgments}

Work was supported by Grant KBN 2P03B 129 14 and partly by Grant KBN PB
1108/P03/95/08. E.Z. acknowledges support from FNRS sabbatical grant at the
University of Liege. She thanks M. Ausloos for stimulating discussions.

\newpage\

\newpage 

{\bf Figure Captions}

\ 

Fig. 1. Persistent currents $I_1$ in a mesoscopic ring as a function of
magnetic flux $\phi /\phi _0$ for different values of disorder. Numerical
results for different $W/t$ are presented by symbols and analytical results
according to formula (\ref{9imp}) for different $\overline{\gamma }$ are
plotted by solid lines.

\ 

Fig. 2. Persistent currents $[I]/I_0$ as a function of magnetic flux $\phi
/\phi _0$ for different parameters $\overline{\gamma }$, in a set of 1-D
mesoscopic rings with even number of conducting electrons $N_e$ in each
ring. Self-sustaining spontaneous currents $I_s$ are denoted by circles.

\ 

Fig. 3. Persistent currents $[I]/I_0$ as a function of magnetic flux $\phi
/\phi _0$ for different temperatures $T$, for the case presented in Fig. 2
with $\overline{\gamma }$ = 20000 \AA .

\ 

Fig. 4. Persistent currents $[I]/I_0$ as a function of magnetic flux $\phi
/\phi _0$ for different parameters $\overline{\gamma }$, in a set of 1-D
mesoscopic rings with odd number of conducting electrons $N_e$ in each ring.
Self-sustaining currents $I_s$ denoted by circles correspond to flux
trapping.

\ 

Fig. 5. Persistent currents $[I]/I_0$ as a function of magnetic flux $\phi
/\phi _0$ for different parameters $\overline{\gamma }$, in a set of 2-D
concentric mesoscopic cylinders with the triangular Fermi surface.
Self-sustaining currents $I_s$ (trapped fluxes) are denoted by circles.

\ 

Fig. 6. Persistent currents $[I]/I_0$ as a function of magnetic flux $\phi
/\phi _0$ for different shapes of the 2-D Fermi surfaces for $\overline{%
\gamma }$ = 50000 \AA . In the inserted figure the shapes of the FS are
shown.

\ 

Fig. 7. Critical fields $H_c$ as a function of the disorder parameter $%
\gamma ^{-1}$ for the cases presented in figures 4, 5.

\ 

Fig. 8. Critical fields $H_c$ as a function of the temperature $T$ for the
cases presented in figures 4, 5.

\end{document}